\documentclass[aps,twocolumn,nopacs,preprintnumbers]{revtex4}

\usepackage{graphicx, fancyhdr, amsfonts, amsmath, amssymb, hyperref, glossaries, siunitx}
\usepackage[T1]{fontenc}
\usepackage{xr}
\newcommand{\subtxt}[2]{\ensuremath{#1_\mathrm{#2}}}

\newcommand{\otwo}[0]{O$_2$}

\newacronym{dqd}{DQD}{double quantum dot}
\newacronym{qd}{QD}{quantum dot}
\newacronym{qw}{QW}{quantum well}
\newacronym{soi}{SOI}{spin--orbit interaction}
\newacronym{hh}{HH}{heavy-hole}
\newacronym{lh}{LH}{light-hole}
\newacronym[\glslongpluralkey={two dimensional hole gases}]{2dhg}{2DHG}{two dimensional hole gas}
\newacronym{haadf}{HAADF}{high-angle annular dark-field}

\newacronym{sige}{SiGe}{silicon-germanium}
\newacronym{ge}{Ge}{germanium}
\newacronym{si}{Si}{silicon}
\newacronym{pt}{Pt}{platinum}
\newacronym{al}{Al}{aluminium}
\newacronym{hf}{HF}{hydrofluoric acid}

\newacronym{sem}{SEM}{scanning electron microscopy}
\newacronym{tem}{TEM}{scanning tunnelling microscopy}
\newacronym{cvd}{CVD}{chemical vapor deposition}
\newacronym{ald}{ALD}{atomic layer depostion}
\newacronym{rie}{RIE}{reactive ion etching}
\newacronym{sdh}{SdH}{Shubnikov–de Haas oscillations}
\newacronym{dhg}{2DHG}{two-dimensional hole gas}
\newacronym{edx}{EDX}{energy dispersive X-Ray}
\newacronym{icp}{ICP}{inductively coupled plasma}
\newacronym{tlf}{TLF}{two-level fluctuators}
\newacronym{flp}{FLP}{fermi-level pinning}
\newacronym{vb}{VB}{valence band}
\newacronym{fl}{E$_F$}{Fermi-level}
\newacronym{alo}{Al$_2$O$_3$}{aluminium oxide}
\newacronym{sin}{SiN$_x$}{silicon nitride}
\newacronym{eb}{E$_b$}{thermal-activation energy}
\newacronym{np}{n$_p$}{percolation density}
\newacronym{cnl}{CNL}{charge neutrality level}
\newacronym{mos}{MOS}{metal-oxide semiconductor}
\newacronym{str}{s}{strained}

\DeclareSIUnit\torr{Torr}
\DeclareSIUnit\sccm{sccm}

\begin{document}

\title{Impact of surface treatments on the transport properties of germanium 2DHGs}

\author{Nikunj Sangwan*}
\author{Eric Jutzi*}
\author{Christian Olsen}
\author{Sarah Vogel}
\author{Arianna Nigro}
\affiliation {\it University of Basel, Klingelbergstrasse 82, 4056 Basel, Switzerland \\
* These two authors contributed equally to this work}
\author{Ilaria Zardo}
\author{Andrea Hofmann}
\affiliation {\it University of Basel, Klingelbergstrasse 82, 4056 Basel, Switzerland}
\affiliation {\it Swiss Nanoscience Institute, Klingelbergstrasse 82, 4056 Basel, Switzerland}

\date{\today}

\begin{abstract} 
Holes in planar \gls{ge} heterostructures show promise for quantum applications, particularly in superconducting and spin qubits, due to strong spin–orbit interaction, low effective mass, and absence of valley degeneracies. However, charge traps cause issues such as gate hysteresis and charge noise. This study examines the effect of surface treatments on the accumulation behaviour and transport properties of \gls{ge}-based \glspl{2dhg}. Oxygen plasma treatment reduces conduction in a setting without applied top-gate voltage and improves the mobility and lowers the percolation density, while \gls{hf} etching provides no benefit. The results suggest that interface traps from the partially oxidised \gls{si} cap pin the Fermi level, and that oxygen plasma reduces the trap density by fully oxidising the Si cap. Therefore, optimising surface treatments is crucial for minimising the charge traps and thereby enhancing the device's performance.

\end{abstract}


\maketitle 

\section{Introduction}
Holes in germanium (Ge) have garnered significant attention for quantum applications, particularly in superconducting \cite{zhuo_hole-type_2023, sagi_gate_2024, zheng_coherent_2024} and spin qubits \cite{watzinger_germanium_2018,jirovec_singlet-triplet_2021,hendrickx_four-qubit_2021,scappucci_germanium_2021}. Their appeal comes from favourable material properties, including strong \gls{soi} \cite{kloeffel_strong_2011,froning_ultrafast_2021}, low effective mass, low hyperfine interaction, and the absence of valley degeneracies \cite{terrazos_theory_2021} and piezoelectricity \cite{scappucci_germanium_2021}. Planar Ge heterostructures are known for their high hole mobility \cite{myronov_holes_2023, lodari_lightly_2022} and low percolation density \cite{lodari_low_2021, nigro_high_2024} and they promise scalable device architectures \cite{borsoi_shared_2024}. The field has rapidly advanced from demonstrating single qubit operations \cite{hendrickx_single-hole_2020, jirovec_singlet-triplet_2021} and simple Josephson junctions \cite{vigneau_germanium_2019,hendrickx_ballistic_2019,aggarwal_enhancement_2021,tosato_hard_2023} to achieving coherent coupling of four qubits \cite{hendrickx_four-qubit_2021} and realising gatemon qubits \cite{sagi_gate_2024}. However, planar Ge heterostructures face challenges due to charge traps distributed throughout the stack. These traps lead to issues such as gate hysteresis \cite{massai_impact_2023,ruggiero_backgate_2024}, which complicates device operation, charge noise that limits coherence \cite{yoneda_quantum-dot_2018, kuhlmann_charge_2013, bermeister_charge_2014, lodari_low_2021,sagi_gate_2024} and two-level fluctuators that induce loss channels for microwave resonators \cite{nigro_demonstration_2024}. Addressing these issues is crucial for improving device reproducibility and performance.

Charge traps are often introduced after the wafer is removed from the growth chamber and during the post-growth fabrication processes, e.g. when clean surfaces are exposed to contaminants and \gls{si} and \gls{ge} form oxides. Cleaning steps, such as \otwo\ plasma treatment or \gls{hf} etching, have been shown to remove organic impurities, dangling bonds, and oxides, creating smoother and cleaner interfaces for \gls{si} and, to some extent, \gls{ge} surfaces \cite{sun_surface_2006, ponath_preparation_2013, ponath_ge001_2017}. Additionally, the gate dielectric, typically \gls{alo}, can host impurities. The quality of the gate dielectric depends on the deposition and annealing conditions \cite{groner_electrical_2002,groner_low-temperature_2004,sioncke_thermal_2009,kim_annealing_2017, kim_influence_2022, paghi_cryogenic_2024}. 

In this study, we systematically investigate the impact of surface treatments, oxide deposition, and oxide annealing conditions on the equilibrium energy levels and the transport behaviour of the \gls{2dhg} in a \gls{ge} \gls{qw}. Even though all our heterostructures are undoped, we observe that some fabrication schemes induce a conducting channel at zero top-gate voltage, while others do not. We propose a straightforward explanation based on the strong Fermi level pinning in Ge. This framework also explains the variations in density tunability, mobility, and percolation density observed in devices subjected to different surface treatments.

\section{Samples}
\begin{figure*}
    \includegraphics[width=\linewidth]{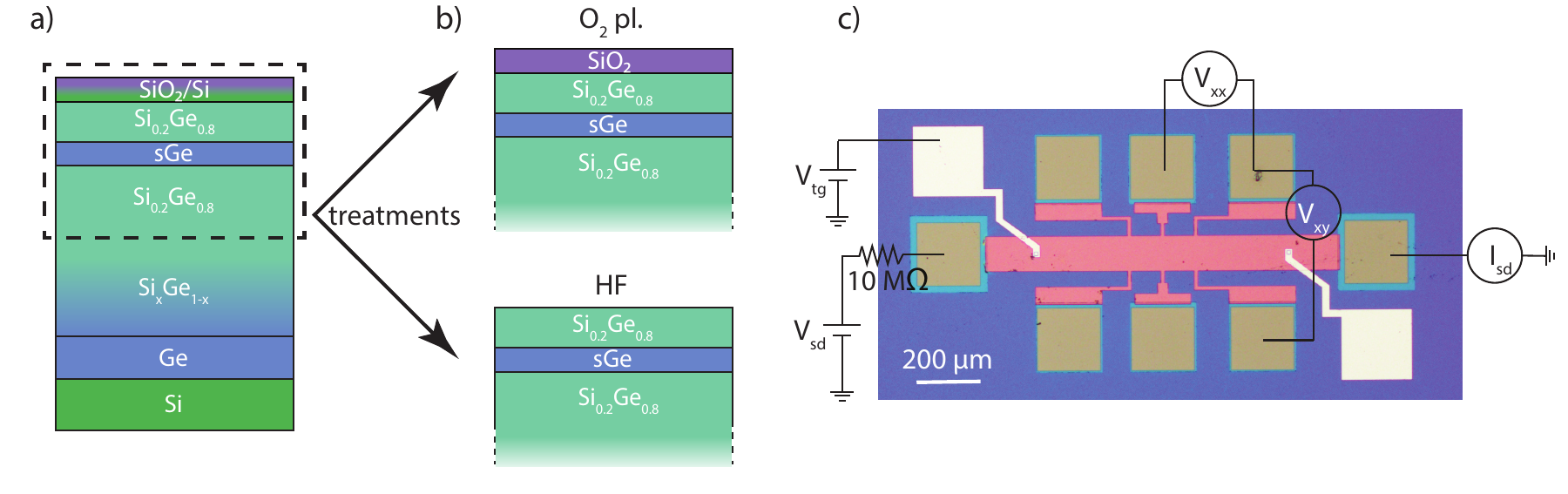}
    \caption{Device schematics: (a) Cross-section of the reverse graded \gls{ge} heterostructure with a strained \gls{ge} (\gls{str}\gls{ge}) \gls{qw}, and (b) the effects of the surface treatments on the \gls{si} cap of the \gls{ge} heterostructure. (c) Measurement schematic of a single-gated layer Hall bar device.}
    \label{fig:Figure1}
\end{figure*}
A cross-sectional schematic of the reverse graded Ge/SiGe heterostructure used in this study is shown in \autoref{fig:Figure1}(a). The heterostructure was grown by chemical vapour deposition as described in Ref.~\cite{nigro_high_2024}, with an additional \SI{5}{\minute} oxidation step in an O$_2$ environment at \SI{500}{\celsius} performed in the growth chamber, unless specified otherwise.

We fabricated two types of devices, a set of simple devices with ohmic contacts and an \gls{alo} layer and another set of devices with an additional top-gate. The fabrication steps are described in Ref.~\cite{nigro_high_2024}. An additional surface treatment was performed during the fabrication process. Specifically, we test the following surface treatments: "O$_{2}$", "\gls{hf}", "O$_{2}$+\gls{hf}" or "as-grown". The devices with surface treatment "O$_{2}$" were oxidised in an O$_{2}$ plasma for \SI{10}{\minute} at \SI{60}{\watt} in an O$_{2}$ flow of \SI{20}{\sccm} before any fabrication step. The devices with surface treatment "\gls{hf}" were dipped in a \SI{2.3}{\percent} \gls{hf} solution for \SI{60}{\second} followed by a \SI{10}{\second} rinse in de-ionised water, after the deposition of ohmic contacts and directly before growing the aluminium oxide. The treatment "O$_{2}$+\gls{hf}" involves the combination of both surface treatments mentioned above, i.e. "O$_{2}$" followed by contact deposition and then "\gls{hf}" before depositing the oxide. No treatment was performed for the "as-grown" devices. Other than described in Ref.~\cite{nigro_high_2024}, our ungated devices do not include any \gls{sin}, and the ohmic contacts were annealed in forming gas for \SI{50}{\minute} at \SI{300}{\celsius}. Due to the strong Fermi level pinning of \gls{ge}, the ohmic contact resistance measured at \SI{4}{\kelvin} typically is as low as $\sim$ \SI{1}{\kilo\ohm}. 

\section{Ungated devices}\label{ungated}
\begin{figure*}
    \includegraphics[width=\linewidth]{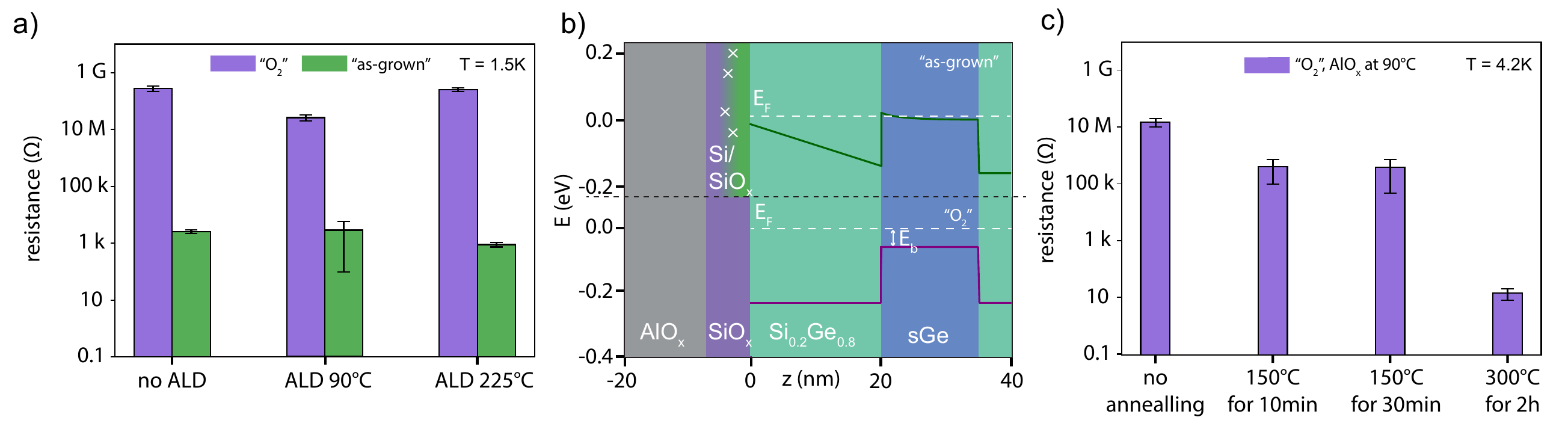}
    \caption{Ungated devices: (a) Two-probe resistance of ohmic contact pairs for different deposition temperatures and treatments "as-grown" and "\otwo". The second and third pair of bars show AlO$_x$ deposited at temperatures of \SI{90}{\celsius} \SI{225}{\celsius}. (b) Representational valence band energy diagrams for two cases. Top: The valence is bent due to the Fermi level pinning to the charge traps. Bottom: Band diagram expected without trap states. (c) Two-probe resistance of ohmic contact pairs for different annealing conditions.}
    \label{fig:Figure2}
\end{figure*}
In the initial set of experiments, we study the simple devices without a top-gate and verify whether charges are accumulated in the undoped \gls{qw} for various device configurations described in the following. We test two surface treatments, "as-grown" and "\otwo" and study the influence of the presence of \gls{alo} and the temperature at which it is grown. For each configuration, we perform two-probe measurements at a temperature of $T=\SI{1.5}{\kelvin}$ of various contact pairs and plot the average resistance in \autoref{fig:Figure2}(a). Additional measurements have been performed to verify that the measured conductance is due to holes in the \gls{qw} rather than any conducting channel on the surface (see supplementary information).
The results in \autoref{fig:Figure2}(a) illustrate that the "\otwo" devices have a higher resistance ($R \geq \SI{15}{\mega\ohm}$) than the "as-grown" devices ($R\sim\SI{1}{\kilo\ohm}$). In contrast, neither the presence of the \gls{alo} nor its deposition temperature dramatically change the resistance of the sample. These results indicate that the mechanism responsible for the measured conduction do not originate in the deposited oxide but in surface-near layers in the wafer.

It is well known that unpassivated \gls{ge} easily forms surface states, leading to a strong Fermi level pinning \cite{dimoulas_fermi-level_2006} if enough charges are available. Typically, the charge neutrality level in Ge \gls{mos} structures amounts to $\sim$ \SI{0.1}{\electronvolt} above the valence band \cite{dimoulas_fermi-level_2006}. The resulting band bending may be strong enough to move the valence band above the Fermi energy for our heterostructures, thereby inducing a \gls{2dhg} without a top-gate. Two possible scenarios are shown in \autoref{fig:Figure2}(b). The top part of the figure displays how the presence of the interface states lead to a strong Fermi level pinning and hence to a significant band bending such that the valence band rises above the Fermi level in the \gls{qw}. On the other hand, the bottom part shows that the valence band remains below the Fermi level in the presence of only few impurities. Nevertheless, the energy difference can be small enough to allow for thermal excitation of charge carriers in the \gls{qw} even at low temperatures. The conduction of the ungated \gls{qw} may therefore be used as an estimate of the amount of charge traps close to the wafer surface. The large resistance measured in the "\otwo" devices indicates that the valence band remains below the Fermi level, while the low resistance in the "as-grown" devices hints to the accumulation of a \gls{2dhg} even in the absence of a top-gate. The large amount of surface states in the latter case likely arises from imperfect oxidation of the cap. Even though \SI{2}{\nano\meter} of Si are expected to completely oxidize in air \cite{morita_growth_1990,bohling_self-limitation_2016}, our amorphous Si cap may remain partially unoxidized due to the growth conditions in which it has been deposited \cite{degli_esposti_wafer-scale_2022} or due to uncertainties in its thickness. In contrast, the oxygen plasma helps to fully oxidize the Si cap and, additionally, it removes any kind of residual polymers. Both effects act to decrease the amount of impurities and thus to a smaller band bending.

Further, we studied the effect of annealing the \gls{alo} on the amount of charge traps. For this study, we used a heterostructure without in-chamber oxidation. We fabricated "\otwo" devices with \gls{alo} grown at \SI{90}{\celsius} and subsequent annealing at different temperatures and times. We perform two-probe measurements at $T=\SI{4}{\kelvin}$ and plot in \autoref{fig:Figure2}(c) the average and standard deviation of the resistance measured for different contact pairs. The resistance decreases with increasing annealing temperature and time, indicating a larger amount of charge traps. It is known that annealing \gls{alo} on Si increases the amount of fixed charges and decreases the amount of interface charges \cite{xu_impact_2021,benick_effect_2010,weber_native_2011,hensling_unraveling_2017}. From our data, we conclude that the combined effect is an increased band bending that induces a \gls{2dhg} in the \gls{qw} when the annealing temperature and time are high enough.

\section{Top-gated devices}
\begin{figure*}
    \centering
    \includegraphics[width=\linewidth]{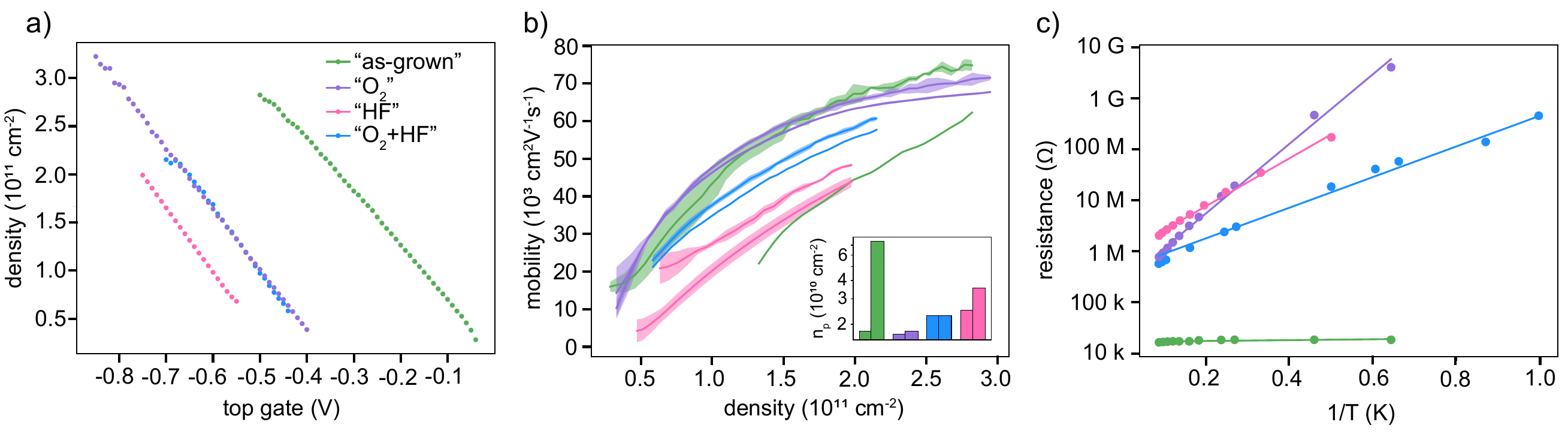}
    \caption{Hall bar devices: (a) Hall densities as a function of top-gate for all the treatments (b) Hall mobilities as a function of Hall density and \gls{np} (inset) for all treatments. The two different data-set per treatment show results from two different devices. The green data-set with the lowest mobilities has no error bars, as for this device we only measured one Hall trace. (c) Two-probe resistance of ohmic contact pair as a function of temperature at zero top gate voltage for different surface treatments.}
    \label{fig:Figure3}
\end{figure*}
We now focus on the effects of different surface treatments on the transport properties such as the \gls{2dhg} density, mobility and percolation density. Hall bar devices as shown in \autoref{fig:Figure1}(c) with all the four mentioned surface treatments were fabricated. Due to the observed depletion-mode behaviour, a mesa was etched for the \otwo\ devices. Standard lock-in magnetotransport measurements were performed to extract the transport properties from the measured longitudinal and Hall resistances at temperatures $T\leq\SI{4}{\kelvin}$, where the mobility \textit{versus} density is independent of temperature (see supplementary information)

We first study the tunability of the density by the top-gate. \autoref{fig:Figure3}(a) shows a linear increase in density as a function of top-gate voltage, as expected for a capacitively coupled top-gate. From the slope, we extract a density tunability of \SI{5.51}{\centi\meter^{-2}/\volt} for the "as-grown" sample and \SI{6.47}{\centi\meter^{-2}/\volt} for the devices with surface treatment. The range of density tunability is limited by a hysteretic shift, which is observed when crossing a certain negative top-gate voltage. The device can only be reset by a thermal cycle to room temperature \cite{ruggiero_backgate_2024} (see supplementary information).
As reported in Ref.~\cite{massai_impact_2023, martinez_variability_2024} the density tunability is severely affected by charge traps at the interface between the sample and the oxide or within the oxide. Charges tunnel from the \gls{2dhg} into these traps and screen the top-gate voltage. The lower density tunability measured in the "as-grown" devices is, therefore, again a signature of the increased amount of charge traps compared to the devices with surface treatment.

Next, we study how the surface treatment influences the mobility and the percolation density. 
In general, for our heterostructures, the mobility is limited by remote scatterers in the low-density regime, and by close-by scatterers in the high-density regime (see supplementary information).
For each surface treatment, the mobility has been measured in different top-gate voltage configurations, i.e. before and after hysteretic shifts, but not after saturation (see supplementary information),
and for at least two devices. The average mobility and standard deviation are plotted in \autoref{fig:Figure3}(b), which show that "\otwo" consistently has the highest mobilities, followed by "\otwo + \gls{hf}" and "\gls{hf}". Meanwhile, the "as-grown" data shows a larger variability within a device as well as in-between the two devices. We conclude that the "\otwo" devices have less charge traps than those treated with \gls{hf}. Knowing that \gls{hf} does not effectively passivate \gls{ge} \cite{sun_surface_2006,ponath_ge001_2017}, we suspect that additional traps are created when the \gls{ge}-rich heterostructure interface is exposed to air while moving it from the \gls{hf}-solution to the oxide deposition chamber. The values extracted for the percolation density, \gls{np}, again corroborate our previous findings: "\otwo" devices have the smallest percolation density hinting at a small number of charge traps, while the \gls{hf} treatment seems to induce more traps and hence a larger \gls{np}. The "as-grown" devices again show a variable performance. This variability indicates that the potential landscape induced by the large number of traps may vary significantly in the heterostructure. Some regions might have a rather homogeneous potential landscape without deep traps, leading to mobilities and percolation densities comparable to the "\otwo" devices. In other regions, the presence of a larger amount and an inhomogeneous distribution of deep traps may cause lower mobilities and higher percolation densities.

We now move on to quantify the amount of band bending induced by the charge traps by estimating the energy difference \subtxt{E}{b} between the valence band and the Fermi energy at zero applied top-gate voltage; see \autoref{fig:Figure2}(b). The two-probe resistance was measured as a function of temperature in a range of $T=\SIrange[]{1}{15}{\kelvin}$ for the four surface treatments. In consistency with the previous data, the "as-grown" device has a low resistance that is independent of the temperature, indicating the accumulation of a \gls{2dhg} already at the lowest temperatures. Meanwhile, the other devices showed an activated behaviour with \subtxt{E}{b} amounting to \SI{1.23}{\milli\electronvolt}, \SI{0.84}{\milli\electronvolt}, and \SI{0.57}{\milli\electronvolt} for the "\otwo", "\gls{hf} and "\otwo + \gls{hf}" devices, respectively.

\section{Conclusion}
In conclusion, our investigation of surface treatments and their effects on planar \gls{ge} heterostructures has provided valuable insights into the sources of charge traps and their impact on the device performance. We suspect that the \gls{si} cap is only partially oxidised and that this leads to a significant number of interface trap states, which degrade the electronic properties of the devices. Although \gls{hf} cleaning is commonly used to remove oxides, it is ineffective in this context as it leaves partially unoxidised \gls{si} intact. Our study indicates that oxygen plasma treatment is more effective, as it fully oxidizes the \gls{si} cap, reducing the number of interface traps and improving the overall quality of the interface. However, when \gls{hf} cleaning is combined with oxygen plasma treatment, the \gls{ge}-rich interface is exposed to air and contaminants, introducing new defects and undermining the benefits of the treatment.

The strong Fermi level pinning in \gls{ge} provides a reasonable explanation for our observation that devices with numerous defects exhibit finite conduction even without a top-gate. Furthermore, the thermally activated behaviour we observed indicates that oxygen plasma treatment results in the fewest defects, leading to the least band bending and reproducibly best device operation. These findings underscore the importance of selecting appropriate surface treatments to minimize charge traps and optimize the performance and reproducibility of \gls{ge}-based quantum devices.

\section{Acknowledgements}
This work was supported as a part of NCCR SPIN, a National Centre of Competence in Research, funded by the Swiss National Science Foundation (grant number 225153), and by the Basel QCQT PhD school. We thank Vera Jo Weibel for her support with the energy-band simulations, Luigi Ruggiero for the Ar-milling tests, Francesco Blanda for inputs with the temperature-dependent measurements, and Christian Schönenberger for discussions throughout the project.

N.S. fabricated the devices with the help of S.V. and recipes developed by E.J. N.S. performed the measurements with the help of S.V. and input of E.J., C.O., and A.H. E.J. performed the initial measurements leading up to the project. N.S. analyzed the data with the input from E.J., C.O., and A.H. A.N. and I.Z. grew the heterostructure. N.S. and A.H. wrote the manuscript with the input of all the authors. The project was conceived by A.H., E.J. and N.S., and supervised by A.H.


\section{Supplementary information}
\subsection{Interface conduction tests}\label{acc sige intf}
Our conclusions of the results obtained with the simple devices relies on the assumption that the measured conduction occurs in the \gls{2dhg} rather than in any potentially conducting oxide interface. Therefore, we fabricated devices with contacts to the SiGe/SiO$_x$ interface but not reaching the \gls{2dhg}. We found no conduction between any of these contacts. Additionally, on the conducting "as-grown" devices that, we performed four-probe magnetotransport measurements and we clearly saw the oscillating behaviour resulting form the Shubnikov-de Haas effect of a high-mobility channel which was, however, not confined into a proper hallbar.

\subsection{Accumulation shift}\label{acc shift}
\begin{figure}[h]
    \includegraphics[width=\linewidth]{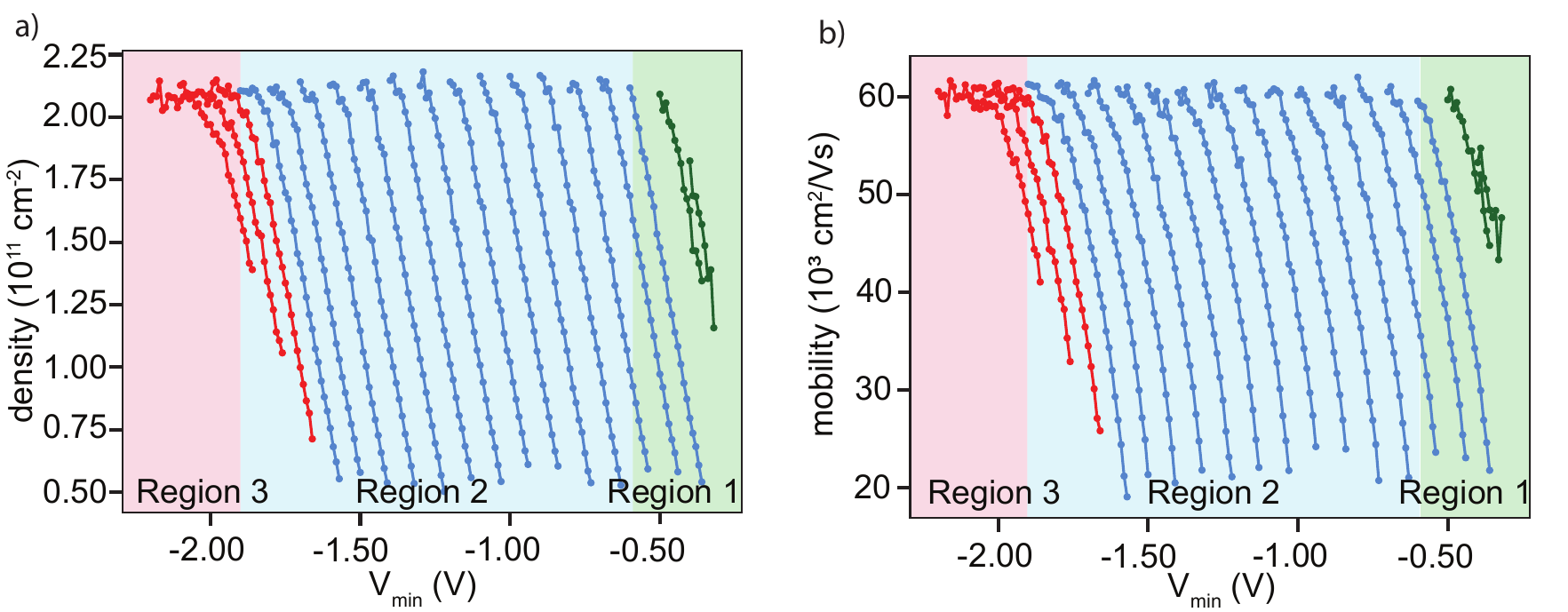}
    \caption{Accumulation shift: Irreversible hysteretic shift of the (a) Hall density, and (b) Hall mobilities with increasingly negative top gate voltage. The measurements shown are obatined with the "\otwo+HF" device.}
    \label{fig:ACC shifts}
\end{figure}
A recent work \cite{massai_impact_2023} reported the observation of hysteretic shifts in gate voltage which could only be reset by thermal cycling. We observe similar shifts, as shown in \autoref{fig:ACC shifts}. The mobility measured at the same density before and after the shift (i.e. in different regimes of top gate voltage) do not match exactly. The accumulation behaviour can be devided into three regions: Region 1: initial accumulation, Region 2: screening regime, positive charges tunneling into the interface states, and Region 3: formation of a triangular-\gls{qw} \cite{massai_impact_2023}. In \autoref{fig:Figure3} of the main text, we report the average mobility and standard deviation measured in the Region 2 for each device.

\subsection{Mobility vs temperature and different scattering regimes}\label{all about mobility}
\begin{figure}[h]
    \includegraphics[width=\linewidth]{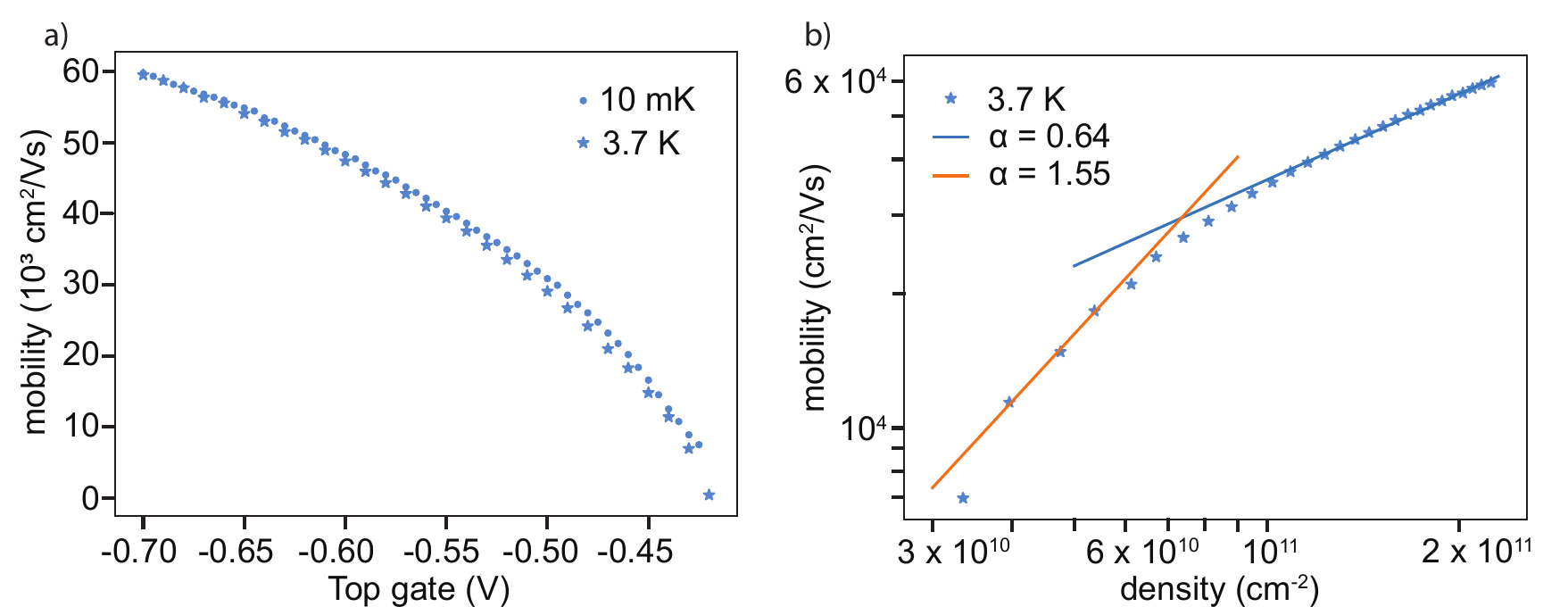}
    \caption{(a) Mobility as a function of top-gate voltage for the "\otwo+HF" Hall bar. (b) Hall mobility as a function of Hall density for the same device with the fits following $\mu \propto n^{\alpha}$}
    \label{fig:scattering regimes}
\end{figure}
\autoref{fig:scattering regimes}(a) shows the mobility of an "\otwo+\gls{hf}" device at two different measurement temperatures, namely \SI{3.7}{\kelvin} and \SI{10}{\milli\kelvin}. The independence of the mobility with respect to these low temperatures indicates that phonon-scattering is not the limiting scattering mechanism \cite{cui_multi-terminal_2015}. \autoref{fig:scattering regimes}(b) shows that at low density, the mobility is limited by scattering at interface states and traps nearby the \gls{qw}. Meanwhile, at high density, it is limited by the background scatterers \cite{monroe_comparison_1993,nigro_high_2024}.

\subsection{Percolation density fits}\label{percolation density fits}
\begin{figure}[h]
    \includegraphics[width=\linewidth]{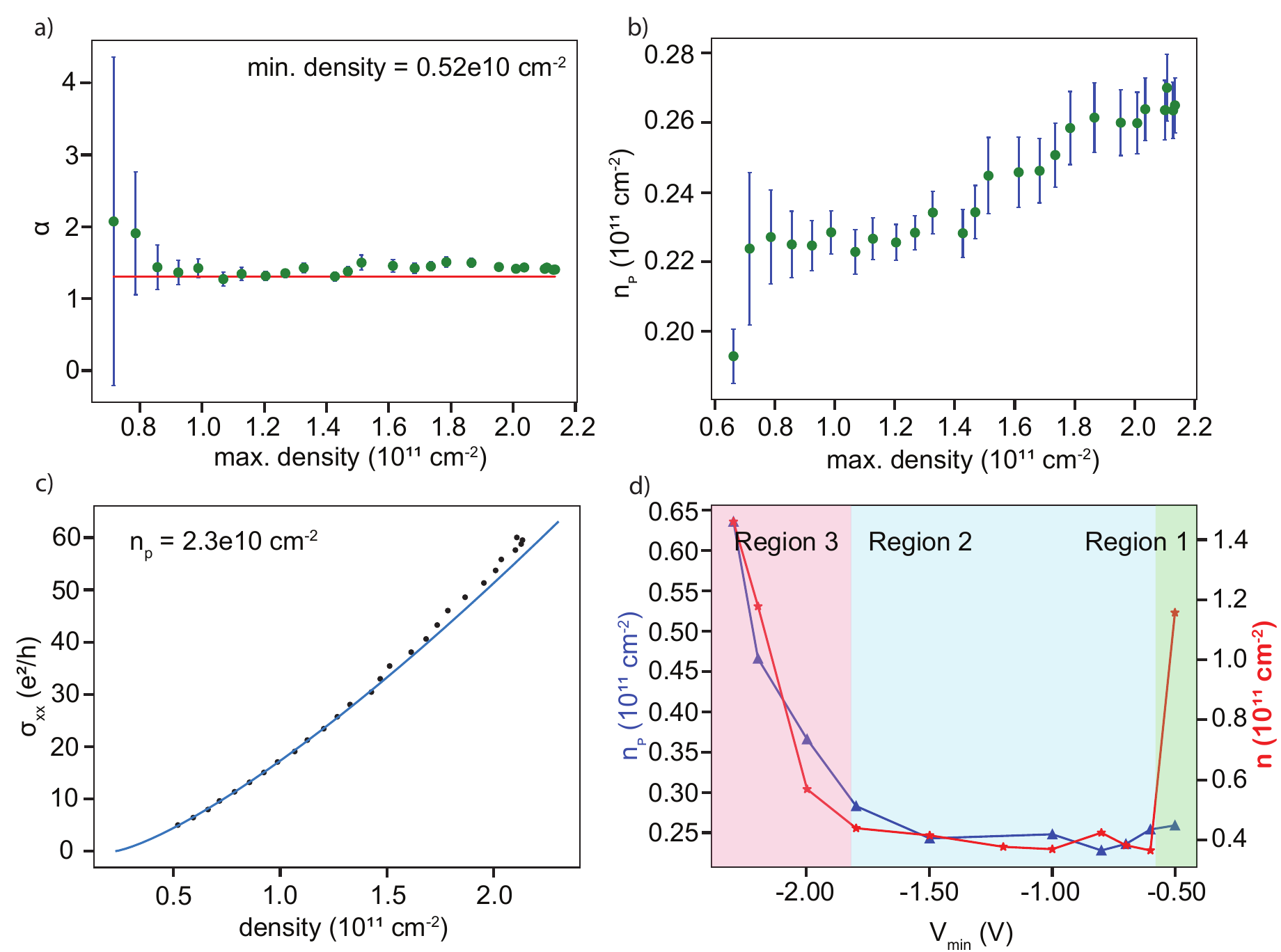}
    \caption{Percolation density fits: Fit parameter (a) $\alpha$, and (b) \gls{np} from fitting $\sigma_{xx} \propto (n-n_{p})^{\alpha}$ and $\sigma_{xx} \propto (n-n_{p})^{1.31}$, respectively, as a function of maximum density used for the fit. (c) Example of fitting \gls{np} using the accurate density range. (d) \gls{np} and measured density at different shifted depletion points for an "\otwo+HF" device.}
    \label{fig:percolation density}
\end{figure}

\gls{np} can be estimated by fitting the conductivity in the low-density regime with the function $\sigma_{xx} \propto (n-n_{p})^{\alpha}$, $\alpha\sim 1.31$ \cite{tracy_observation_2009}, which is valid in the low-density regime. For fitting \gls{np}, this low-density regime first has to be determined. We use the function mentioned above to fit our data with the following procedure: we first fit our data in two different ways, once by using two fit parameters ($\alpha$ and \gls{np}) and once with only one fit parameter (\gls{np}, while fixing $\alpha=1.31$). Both these fits are performed with an increasing range of density used as input values. Examples are shown in \autoref{fig:percolation density}(a) and (b), respectively. Second, we select the density range such that $\alpha=1.31$ and the error bars of the \gls{np} fit with $\alpha=1.31$ remain constant. We then use the selected regime to fit $\sigma_{xx} \propto (n-n_{p})^{\alpha}$ with \gls{np} as the only fit parameter and fixed $\alpha=1.31$.
Extracting percolation density for all the shifted accumulation curves suggests the same three regions as mentioned in \autoref{acc shift}.

\bibliography{surfacetreatments}

\begin{thebibliography}{47}
\expandafter\ifx\csname natexlab\endcsname\relax\def\natexlab#1{#1}\fi
\expandafter\ifx\csname bibnamefont\endcsname\relax
  \def\bibnamefont#1{#1}\fi
\expandafter\ifx\csname bibfnamefont\endcsname\relax
  \def\bibfnamefont#1{#1}\fi
\expandafter\ifx\csname citenamefont\endcsname\relax
  \def\citenamefont#1{#1}\fi
\expandafter\ifx\csname url\endcsname\relax
  \def\url#1{\texttt{#1}}\fi
\expandafter\ifx\csname urlprefix\endcsname\relax\def\urlprefix{URL }\fi
\providecommand{\bibinfo}[2]{#2}
\providecommand{\eprint}[2][]{\url{#2}}

\bibitem[{\citenamefont{Zhuo et~al.}(2023)\citenamefont{Zhuo, Lyu, Sun, Li, Li, Ji, Fan, Bakkers, Han, Song et~al.}}]{zhuo_hole-type_2023}
\bibinfo{author}{\bibfnamefont{E.}~\bibnamefont{Zhuo}}, \bibinfo{author}{\bibfnamefont{Z.}~\bibnamefont{Lyu}}, \bibinfo{author}{\bibfnamefont{X.}~\bibnamefont{Sun}}, \bibinfo{author}{\bibfnamefont{A.}~\bibnamefont{Li}}, \bibinfo{author}{\bibfnamefont{B.}~\bibnamefont{Li}}, \bibinfo{author}{\bibfnamefont{Z.}~\bibnamefont{Ji}}, \bibinfo{author}{\bibfnamefont{J.}~\bibnamefont{Fan}}, \bibinfo{author}{\bibfnamefont{E.~P. a.~M.} \bibnamefont{Bakkers}}, \bibinfo{author}{\bibfnamefont{X.}~\bibnamefont{Han}}, \bibinfo{author}{\bibfnamefont{X.}~\bibnamefont{Song}}, \bibnamefont{et~al.}, \bibinfo{journal}{npj Quantum Information} \textbf{\bibinfo{volume}{9}}, \bibinfo{pages}{1} (\bibinfo{year}{2023}), ISSN \bibinfo{issn}{2056-6387}, \urlprefix\url{https://www.nature.com/articles/s41534-023-00721-9}.

\bibitem[{\citenamefont{Sagi et~al.}(2024)\citenamefont{Sagi, Crippa, Valentini, Janik, Baghumyan, Fabris, Kapoor, Hassani, Fink, Calcaterra et~al.}}]{sagi_gate_2024}
\bibinfo{author}{\bibfnamefont{O.}~\bibnamefont{Sagi}}, \bibinfo{author}{\bibfnamefont{A.}~\bibnamefont{Crippa}}, \bibinfo{author}{\bibfnamefont{M.}~\bibnamefont{Valentini}}, \bibinfo{author}{\bibfnamefont{M.}~\bibnamefont{Janik}}, \bibinfo{author}{\bibfnamefont{L.}~\bibnamefont{Baghumyan}}, \bibinfo{author}{\bibfnamefont{G.}~\bibnamefont{Fabris}}, \bibinfo{author}{\bibfnamefont{L.}~\bibnamefont{Kapoor}}, \bibinfo{author}{\bibfnamefont{F.}~\bibnamefont{Hassani}}, \bibinfo{author}{\bibfnamefont{J.}~\bibnamefont{Fink}}, \bibinfo{author}{\bibfnamefont{S.}~\bibnamefont{Calcaterra}}, \bibnamefont{et~al.}, \bibinfo{journal}{Nature Communications} \textbf{\bibinfo{volume}{15}}, \bibinfo{pages}{6400} (\bibinfo{year}{2024}), ISSN \bibinfo{issn}{2041-1723}, \urlprefix\url{https://www.nature.com/articles/s41467-024-50763-6}.

\bibitem[{\citenamefont{Zheng et~al.}(2024)\citenamefont{Zheng, Cheung, Sangwan, Kononov, Haller, Ridderbos, Ciaccia, Ungerer, Li, Bakkers et~al.}}]{zheng_coherent_2024}
\bibinfo{author}{\bibfnamefont{H.}~\bibnamefont{Zheng}}, \bibinfo{author}{\bibfnamefont{L.~Y.} \bibnamefont{Cheung}}, \bibinfo{author}{\bibfnamefont{N.}~\bibnamefont{Sangwan}}, \bibinfo{author}{\bibfnamefont{A.}~\bibnamefont{Kononov}}, \bibinfo{author}{\bibfnamefont{R.}~\bibnamefont{Haller}}, \bibinfo{author}{\bibfnamefont{J.}~\bibnamefont{Ridderbos}}, \bibinfo{author}{\bibfnamefont{C.}~\bibnamefont{Ciaccia}}, \bibinfo{author}{\bibfnamefont{J.~H.} \bibnamefont{Ungerer}}, \bibinfo{author}{\bibfnamefont{A.}~\bibnamefont{Li}}, \bibinfo{author}{\bibfnamefont{E.~P.} \bibnamefont{Bakkers}}, \bibnamefont{et~al.}, \bibinfo{journal}{Nano Letters} \textbf{\bibinfo{volume}{24}}, \bibinfo{pages}{7173} (\bibinfo{year}{2024}), ISSN \bibinfo{issn}{1530-6984, 1530-6992}, \urlprefix\url{https://pubs.acs.org/doi/10.1021/acs.nanolett.4c00770}.

\bibitem[{\citenamefont{Watzinger et~al.}(2018)\citenamefont{Watzinger, Kukučka, Vukušić, Gao, Wang, Schäffler, Zhang, and Katsaros}}]{watzinger_germanium_2018}
\bibinfo{author}{\bibfnamefont{H.}~\bibnamefont{Watzinger}}, \bibinfo{author}{\bibfnamefont{J.}~\bibnamefont{Kukučka}}, \bibinfo{author}{\bibfnamefont{L.}~\bibnamefont{Vukušić}}, \bibinfo{author}{\bibfnamefont{F.}~\bibnamefont{Gao}}, \bibinfo{author}{\bibfnamefont{T.}~\bibnamefont{Wang}}, \bibinfo{author}{\bibfnamefont{F.}~\bibnamefont{Schäffler}}, \bibinfo{author}{\bibfnamefont{J.-J.} \bibnamefont{Zhang}}, \bibnamefont{and} \bibinfo{author}{\bibfnamefont{G.}~\bibnamefont{Katsaros}}, \bibinfo{journal}{Nature Communications} \textbf{\bibinfo{volume}{9}}, \bibinfo{pages}{3902} (\bibinfo{year}{2018}), ISSN \bibinfo{issn}{2041-1723}, \urlprefix\url{https://www.nature.com/articles/s41467-018-06418-4}.

\bibitem[{\citenamefont{Jirovec et~al.}(2021)\citenamefont{Jirovec, Hofmann, Ballabio, Mutter, Tavani, Botifoll, Crippa, Kukucka, Sagi, Martins et~al.}}]{jirovec_singlet-triplet_2021}
\bibinfo{author}{\bibfnamefont{D.}~\bibnamefont{Jirovec}}, \bibinfo{author}{\bibfnamefont{A.}~\bibnamefont{Hofmann}}, \bibinfo{author}{\bibfnamefont{A.}~\bibnamefont{Ballabio}}, \bibinfo{author}{\bibfnamefont{P.~M.} \bibnamefont{Mutter}}, \bibinfo{author}{\bibfnamefont{G.}~\bibnamefont{Tavani}}, \bibinfo{author}{\bibfnamefont{M.}~\bibnamefont{Botifoll}}, \bibinfo{author}{\bibfnamefont{A.}~\bibnamefont{Crippa}}, \bibinfo{author}{\bibfnamefont{J.}~\bibnamefont{Kukucka}}, \bibinfo{author}{\bibfnamefont{O.}~\bibnamefont{Sagi}}, \bibinfo{author}{\bibfnamefont{F.}~\bibnamefont{Martins}}, \bibnamefont{et~al.}, \bibinfo{journal}{Nature Materials} \textbf{\bibinfo{volume}{20}}, \bibinfo{pages}{1106} (\bibinfo{year}{2021}), ISSN \bibinfo{issn}{1476-4660}, \urlprefix\url{https://www.nature.com/articles/s41563-021-01022-2}.

\bibitem[{\citenamefont{Hendrickx et~al.}(2021)\citenamefont{Hendrickx, Lawrie, Russ, van Riggelen, de~Snoo, Schouten, Sammak, Scappucci, and Veldhorst}}]{hendrickx_four-qubit_2021}
\bibinfo{author}{\bibfnamefont{N.~W.} \bibnamefont{Hendrickx}}, \bibinfo{author}{\bibfnamefont{W.~I.~L.} \bibnamefont{Lawrie}}, \bibinfo{author}{\bibfnamefont{M.}~\bibnamefont{Russ}}, \bibinfo{author}{\bibfnamefont{F.}~\bibnamefont{van Riggelen}}, \bibinfo{author}{\bibfnamefont{S.~L.} \bibnamefont{de~Snoo}}, \bibinfo{author}{\bibfnamefont{R.~N.} \bibnamefont{Schouten}}, \bibinfo{author}{\bibfnamefont{A.}~\bibnamefont{Sammak}}, \bibinfo{author}{\bibfnamefont{G.}~\bibnamefont{Scappucci}}, \bibnamefont{and} \bibinfo{author}{\bibfnamefont{M.}~\bibnamefont{Veldhorst}}, \bibinfo{journal}{Nature} \textbf{\bibinfo{volume}{591}}, \bibinfo{pages}{580} (\bibinfo{year}{2021}), ISSN \bibinfo{issn}{1476-4687}, \urlprefix\url{https://www.nature.com/articles/s41586-021-03332-6}.

\bibitem[{\citenamefont{Scappucci et~al.}(2021)\citenamefont{Scappucci, Kloeffel, Zwanenburg, Loss, Myronov, Zhang, De~Franceschi, Katsaros, and Veldhorst}}]{scappucci_germanium_2021}
\bibinfo{author}{\bibfnamefont{G.}~\bibnamefont{Scappucci}}, \bibinfo{author}{\bibfnamefont{C.}~\bibnamefont{Kloeffel}}, \bibinfo{author}{\bibfnamefont{F.~A.} \bibnamefont{Zwanenburg}}, \bibinfo{author}{\bibfnamefont{D.}~\bibnamefont{Loss}}, \bibinfo{author}{\bibfnamefont{M.}~\bibnamefont{Myronov}}, \bibinfo{author}{\bibfnamefont{J.-J.} \bibnamefont{Zhang}}, \bibinfo{author}{\bibfnamefont{S.}~\bibnamefont{De~Franceschi}}, \bibinfo{author}{\bibfnamefont{G.}~\bibnamefont{Katsaros}}, \bibnamefont{and} \bibinfo{author}{\bibfnamefont{M.}~\bibnamefont{Veldhorst}}, \bibinfo{journal}{Nature Reviews Materials} \textbf{\bibinfo{volume}{6}}, \bibinfo{pages}{926} (\bibinfo{year}{2021}), ISSN \bibinfo{issn}{2058-8437}, \urlprefix\url{https://www.nature.com/articles/s41578-020-00262-z}.

\bibitem[{\citenamefont{Kloeffel et~al.}(2011)\citenamefont{Kloeffel, Trif, and Loss}}]{kloeffel_strong_2011}
\bibinfo{author}{\bibfnamefont{C.}~\bibnamefont{Kloeffel}}, \bibinfo{author}{\bibfnamefont{M.}~\bibnamefont{Trif}}, \bibnamefont{and} \bibinfo{author}{\bibfnamefont{D.}~\bibnamefont{Loss}}, \bibinfo{journal}{Physical Review B} \textbf{\bibinfo{volume}{84}}, \bibinfo{pages}{195314} (\bibinfo{year}{2011}), \urlprefix\url{https://link.aps.org/doi/10.1103/PhysRevB.84.195314}.

\bibitem[{\citenamefont{Froning et~al.}(2021)\citenamefont{Froning, Camenzind, van~der Molen, Li, Bakkers, Zumbühl, and Braakman}}]{froning_ultrafast_2021}
\bibinfo{author}{\bibfnamefont{F.~N.~M.} \bibnamefont{Froning}}, \bibinfo{author}{\bibfnamefont{L.~C.} \bibnamefont{Camenzind}}, \bibinfo{author}{\bibfnamefont{O.~A.~H.} \bibnamefont{van~der Molen}}, \bibinfo{author}{\bibfnamefont{A.}~\bibnamefont{Li}}, \bibinfo{author}{\bibfnamefont{E.~P. A.~M.} \bibnamefont{Bakkers}}, \bibinfo{author}{\bibfnamefont{D.~M.} \bibnamefont{Zumbühl}}, \bibnamefont{and} \bibinfo{author}{\bibfnamefont{F.~R.} \bibnamefont{Braakman}}, \bibinfo{journal}{Nature Nanotechnology} \textbf{\bibinfo{volume}{16}}, \bibinfo{pages}{308} (\bibinfo{year}{2021}), ISSN \bibinfo{issn}{1748-3395}, \urlprefix\url{https://www.nature.com/articles/s41565-020-00828-6}.

\bibitem[{\citenamefont{Terrazos et~al.}(2021)\citenamefont{Terrazos, Marcellina, Wang, Coppersmith, Friesen, Hamilton, Hu, Koiller, Saraiva, Culcer et~al.}}]{terrazos_theory_2021}
\bibinfo{author}{\bibfnamefont{L.~A.} \bibnamefont{Terrazos}}, \bibinfo{author}{\bibfnamefont{E.}~\bibnamefont{Marcellina}}, \bibinfo{author}{\bibfnamefont{Z.}~\bibnamefont{Wang}}, \bibinfo{author}{\bibfnamefont{S.~N.} \bibnamefont{Coppersmith}}, \bibinfo{author}{\bibfnamefont{M.}~\bibnamefont{Friesen}}, \bibinfo{author}{\bibfnamefont{A.~R.} \bibnamefont{Hamilton}}, \bibinfo{author}{\bibfnamefont{X.}~\bibnamefont{Hu}}, \bibinfo{author}{\bibfnamefont{B.}~\bibnamefont{Koiller}}, \bibinfo{author}{\bibfnamefont{A.~L.} \bibnamefont{Saraiva}}, \bibinfo{author}{\bibfnamefont{D.}~\bibnamefont{Culcer}}, \bibnamefont{et~al.}, \bibinfo{journal}{Physical Review B} \textbf{\bibinfo{volume}{103}}, \bibinfo{pages}{125201} (\bibinfo{year}{2021}), ISSN \bibinfo{issn}{2469-9950, 2469-9969}, \urlprefix\url{https://link.aps.org/doi/10.1103/PhysRevB.103.125201}.

\bibitem[{\citenamefont{Myronov et~al.}(2023)\citenamefont{Myronov, Kycia, Waldron, Jiang, Barrios, Bogan, Coleridge, and Studenikin}}]{myronov_holes_2023}
\bibinfo{author}{\bibfnamefont{M.}~\bibnamefont{Myronov}}, \bibinfo{author}{\bibfnamefont{J.}~\bibnamefont{Kycia}}, \bibinfo{author}{\bibfnamefont{P.}~\bibnamefont{Waldron}}, \bibinfo{author}{\bibfnamefont{W.}~\bibnamefont{Jiang}}, \bibinfo{author}{\bibfnamefont{P.}~\bibnamefont{Barrios}}, \bibinfo{author}{\bibfnamefont{A.}~\bibnamefont{Bogan}}, \bibinfo{author}{\bibfnamefont{P.}~\bibnamefont{Coleridge}}, \bibnamefont{and} \bibinfo{author}{\bibfnamefont{S.}~\bibnamefont{Studenikin}}, \bibinfo{journal}{Small Science} \textbf{\bibinfo{volume}{3}}, \bibinfo{pages}{2200094} (\bibinfo{year}{2023}), ISSN \bibinfo{issn}{2688-4046}.

\bibitem[{\citenamefont{Lodari et~al.}(2022)\citenamefont{Lodari, Kong, Rendell, Tosato, Sammak, Veldhorst, Hamilton, and Scappucci}}]{lodari_lightly_2022}
\bibinfo{author}{\bibfnamefont{M.}~\bibnamefont{Lodari}}, \bibinfo{author}{\bibfnamefont{O.}~\bibnamefont{Kong}}, \bibinfo{author}{\bibfnamefont{M.}~\bibnamefont{Rendell}}, \bibinfo{author}{\bibfnamefont{A.}~\bibnamefont{Tosato}}, \bibinfo{author}{\bibfnamefont{A.}~\bibnamefont{Sammak}}, \bibinfo{author}{\bibfnamefont{M.}~\bibnamefont{Veldhorst}}, \bibinfo{author}{\bibfnamefont{A.~R.} \bibnamefont{Hamilton}}, \bibnamefont{and} \bibinfo{author}{\bibfnamefont{G.}~\bibnamefont{Scappucci}}, \bibinfo{journal}{Applied Physics Letters} \textbf{\bibinfo{volume}{120}}, \bibinfo{pages}{122104} (\bibinfo{year}{2022}), ISSN \bibinfo{issn}{0003-6951, 1077-3118}, \urlprefix\url{https://pubs.aip.org/apl/article/120/12/122104/2833198/Lightly-strained-germanium-quantum-wells-with-hole}.

\bibitem[{\citenamefont{Lodari et~al.}(2021)\citenamefont{Lodari, Hendrickx, Lawrie, Hsiao, Vandersypen, Sammak, Veldhorst, and Scappucci}}]{lodari_low_2021}
\bibinfo{author}{\bibfnamefont{M.}~\bibnamefont{Lodari}}, \bibinfo{author}{\bibfnamefont{N.~W.} \bibnamefont{Hendrickx}}, \bibinfo{author}{\bibfnamefont{W.~I.~L.} \bibnamefont{Lawrie}}, \bibinfo{author}{\bibfnamefont{T.-K.} \bibnamefont{Hsiao}}, \bibinfo{author}{\bibfnamefont{L.~M.~K.} \bibnamefont{Vandersypen}}, \bibinfo{author}{\bibfnamefont{A.}~\bibnamefont{Sammak}}, \bibinfo{author}{\bibfnamefont{M.}~\bibnamefont{Veldhorst}}, \bibnamefont{and} \bibinfo{author}{\bibfnamefont{G.}~\bibnamefont{Scappucci}}, \bibinfo{journal}{Materials for Quantum Technology} \textbf{\bibinfo{volume}{1}}, \bibinfo{pages}{011002} (\bibinfo{year}{2021}), ISSN \bibinfo{issn}{2633-4356}, \urlprefix\url{https://iopscience.iop.org/article/10.1088/2633-4356/abcd82}.

\bibitem[{\citenamefont{Nigro et~al.}(2024{\natexlab{a}})\citenamefont{Nigro, Jutzi, Forrer, Hofmann, Gadea, and Zardo}}]{nigro_high_2024}
\bibinfo{author}{\bibfnamefont{A.}~\bibnamefont{Nigro}}, \bibinfo{author}{\bibfnamefont{E.}~\bibnamefont{Jutzi}}, \bibinfo{author}{\bibfnamefont{N.}~\bibnamefont{Forrer}}, \bibinfo{author}{\bibfnamefont{A.}~\bibnamefont{Hofmann}}, \bibinfo{author}{\bibfnamefont{G.}~\bibnamefont{Gadea}}, \bibnamefont{and} \bibinfo{author}{\bibfnamefont{I.}~\bibnamefont{Zardo}}, \bibinfo{journal}{Physical Review Materials} \textbf{\bibinfo{volume}{8}}, \bibinfo{pages}{066201} (\bibinfo{year}{2024}{\natexlab{a}}), \urlprefix\url{https://link.aps.org/doi/10.1103/PhysRevMaterials.8.066201}.

\bibitem[{\citenamefont{Borsoi et~al.}(2024)\citenamefont{Borsoi, Hendrickx, John, Meyer, Motz, van Riggelen, Sammak, de~Snoo, Scappucci, and Veldhorst}}]{borsoi_shared_2024}
\bibinfo{author}{\bibfnamefont{F.}~\bibnamefont{Borsoi}}, \bibinfo{author}{\bibfnamefont{N.~W.} \bibnamefont{Hendrickx}}, \bibinfo{author}{\bibfnamefont{V.}~\bibnamefont{John}}, \bibinfo{author}{\bibfnamefont{M.}~\bibnamefont{Meyer}}, \bibinfo{author}{\bibfnamefont{S.}~\bibnamefont{Motz}}, \bibinfo{author}{\bibfnamefont{F.}~\bibnamefont{van Riggelen}}, \bibinfo{author}{\bibfnamefont{A.}~\bibnamefont{Sammak}}, \bibinfo{author}{\bibfnamefont{S.~L.} \bibnamefont{de~Snoo}}, \bibinfo{author}{\bibfnamefont{G.}~\bibnamefont{Scappucci}}, \bibnamefont{and} \bibinfo{author}{\bibfnamefont{M.}~\bibnamefont{Veldhorst}}, \bibinfo{journal}{Nature Nanotechnology} \textbf{\bibinfo{volume}{19}}, \bibinfo{pages}{21} (\bibinfo{year}{2024}), ISSN \bibinfo{issn}{1748-3395}, \urlprefix\url{https://www.nature.com/articles/s41565-023-01491-3}.

\bibitem[{\citenamefont{Hendrickx et~al.}(2020)\citenamefont{Hendrickx, Lawrie, Petit, Sammak, Scappucci, and Veldhorst}}]{hendrickx_single-hole_2020}
\bibinfo{author}{\bibfnamefont{N.~W.} \bibnamefont{Hendrickx}}, \bibinfo{author}{\bibfnamefont{W.~I.~L.} \bibnamefont{Lawrie}}, \bibinfo{author}{\bibfnamefont{L.}~\bibnamefont{Petit}}, \bibinfo{author}{\bibfnamefont{A.}~\bibnamefont{Sammak}}, \bibinfo{author}{\bibfnamefont{G.}~\bibnamefont{Scappucci}}, \bibnamefont{and} \bibinfo{author}{\bibfnamefont{M.}~\bibnamefont{Veldhorst}}, \bibinfo{journal}{Nature Communications} \textbf{\bibinfo{volume}{11}}, \bibinfo{pages}{3478} (\bibinfo{year}{2020}), ISSN \bibinfo{issn}{2041-1723}, \urlprefix\url{https://www.nature.com/articles/s41467-020-17211-7}.

\bibitem[{\citenamefont{Vigneau et~al.}(2019)\citenamefont{Vigneau, Mizokuchi, Zanuz, Huang, Tan, Maurand, Frolov, Sammak, Scappucci, Lefloch et~al.}}]{vigneau_germanium_2019}
\bibinfo{author}{\bibfnamefont{F.}~\bibnamefont{Vigneau}}, \bibinfo{author}{\bibfnamefont{R.}~\bibnamefont{Mizokuchi}}, \bibinfo{author}{\bibfnamefont{D.~C.} \bibnamefont{Zanuz}}, \bibinfo{author}{\bibfnamefont{X.}~\bibnamefont{Huang}}, \bibinfo{author}{\bibfnamefont{S.}~\bibnamefont{Tan}}, \bibinfo{author}{\bibfnamefont{R.}~\bibnamefont{Maurand}}, \bibinfo{author}{\bibfnamefont{S.}~\bibnamefont{Frolov}}, \bibinfo{author}{\bibfnamefont{A.}~\bibnamefont{Sammak}}, \bibinfo{author}{\bibfnamefont{G.}~\bibnamefont{Scappucci}}, \bibinfo{author}{\bibfnamefont{F.}~\bibnamefont{Lefloch}}, \bibnamefont{et~al.}, \bibinfo{journal}{Nano Letters} \textbf{\bibinfo{volume}{19}}, \bibinfo{pages}{1023} (\bibinfo{year}{2019}), ISSN \bibinfo{issn}{1530-6984, 1530-6992}, \urlprefix\url{https://pubs.acs.org/doi/10.1021/acs.nanolett.8b04275}.

\bibitem[{\citenamefont{Hendrickx et~al.}(2019)\citenamefont{Hendrickx, Tagliaferri, Kouwenhoven, Li, Franke, Sammak, Brinkman, Scappucci, and Veldhorst}}]{hendrickx_ballistic_2019}
\bibinfo{author}{\bibfnamefont{N.~W.} \bibnamefont{Hendrickx}}, \bibinfo{author}{\bibfnamefont{M.~L.~V.} \bibnamefont{Tagliaferri}}, \bibinfo{author}{\bibfnamefont{M.}~\bibnamefont{Kouwenhoven}}, \bibinfo{author}{\bibfnamefont{R.}~\bibnamefont{Li}}, \bibinfo{author}{\bibfnamefont{D.~P.} \bibnamefont{Franke}}, \bibinfo{author}{\bibfnamefont{A.}~\bibnamefont{Sammak}}, \bibinfo{author}{\bibfnamefont{A.}~\bibnamefont{Brinkman}}, \bibinfo{author}{\bibfnamefont{G.}~\bibnamefont{Scappucci}}, \bibnamefont{and} \bibinfo{author}{\bibfnamefont{M.}~\bibnamefont{Veldhorst}}, \bibinfo{journal}{Physical Review B} \textbf{\bibinfo{volume}{99}}, \bibinfo{pages}{075435} (\bibinfo{year}{2019}), ISSN \bibinfo{issn}{2469-9950, 2469-9969}, \urlprefix\url{https://link.aps.org/doi/10.1103/PhysRevB.99.075435}.

\bibitem[{\citenamefont{Aggarwal et~al.}(2021)\citenamefont{Aggarwal, Hofmann, Jirovec, Prieto, Sammak, Botifoll, Martí-Sánchez, Veldhorst, Arbiol, Scappucci et~al.}}]{aggarwal_enhancement_2021}
\bibinfo{author}{\bibfnamefont{K.}~\bibnamefont{Aggarwal}}, \bibinfo{author}{\bibfnamefont{A.}~\bibnamefont{Hofmann}}, \bibinfo{author}{\bibfnamefont{D.}~\bibnamefont{Jirovec}}, \bibinfo{author}{\bibfnamefont{I.}~\bibnamefont{Prieto}}, \bibinfo{author}{\bibfnamefont{A.}~\bibnamefont{Sammak}}, \bibinfo{author}{\bibfnamefont{M.}~\bibnamefont{Botifoll}}, \bibinfo{author}{\bibfnamefont{S.}~\bibnamefont{Martí-Sánchez}}, \bibinfo{author}{\bibfnamefont{M.}~\bibnamefont{Veldhorst}}, \bibinfo{author}{\bibfnamefont{J.}~\bibnamefont{Arbiol}}, \bibinfo{author}{\bibfnamefont{G.}~\bibnamefont{Scappucci}}, \bibnamefont{et~al.}, \bibinfo{journal}{Physical Review Research} \textbf{\bibinfo{volume}{3}}, \bibinfo{pages}{L022005} (\bibinfo{year}{2021}), ISSN \bibinfo{issn}{2643-1564}, \urlprefix\url{https://link.aps.org/doi/10.1103/PhysRevResearch.3.L022005}.

\bibitem[{\citenamefont{Tosato et~al.}(2023)\citenamefont{Tosato, Levajac, Wang, Boor, Borsoi, Botifoll, Borja, Martí-Sánchez, Arbiol, Sammak et~al.}}]{tosato_hard_2023}
\bibinfo{author}{\bibfnamefont{A.}~\bibnamefont{Tosato}}, \bibinfo{author}{\bibfnamefont{V.}~\bibnamefont{Levajac}}, \bibinfo{author}{\bibfnamefont{J.-Y.} \bibnamefont{Wang}}, \bibinfo{author}{\bibfnamefont{C.~J.} \bibnamefont{Boor}}, \bibinfo{author}{\bibfnamefont{F.}~\bibnamefont{Borsoi}}, \bibinfo{author}{\bibfnamefont{M.}~\bibnamefont{Botifoll}}, \bibinfo{author}{\bibfnamefont{C.~N.} \bibnamefont{Borja}}, \bibinfo{author}{\bibfnamefont{S.}~\bibnamefont{Martí-Sánchez}}, \bibinfo{author}{\bibfnamefont{J.}~\bibnamefont{Arbiol}}, \bibinfo{author}{\bibfnamefont{A.}~\bibnamefont{Sammak}}, \bibnamefont{et~al.}, \bibinfo{journal}{Communications Materials} \textbf{\bibinfo{volume}{4}}, \bibinfo{pages}{1} (\bibinfo{year}{2023}), ISSN \bibinfo{issn}{2662-4443}, \urlprefix\url{https://www.nature.com/articles/s43246-023-00351-w}.

\bibitem[{\citenamefont{Massai et~al.}(2023)\citenamefont{Massai, Hetényi, Mergenthaler, Schupp, Sommer, Paredes, Bedell, Harvey-Collard, Salis, Fuhrer et~al.}}]{massai_impact_2023}
\bibinfo{author}{\bibfnamefont{L.}~\bibnamefont{Massai}}, \bibinfo{author}{\bibfnamefont{B.}~\bibnamefont{Hetényi}}, \bibinfo{author}{\bibfnamefont{M.}~\bibnamefont{Mergenthaler}}, \bibinfo{author}{\bibfnamefont{F.~J.} \bibnamefont{Schupp}}, \bibinfo{author}{\bibfnamefont{L.}~\bibnamefont{Sommer}}, \bibinfo{author}{\bibfnamefont{S.}~\bibnamefont{Paredes}}, \bibinfo{author}{\bibfnamefont{S.~W.} \bibnamefont{Bedell}}, \bibinfo{author}{\bibfnamefont{P.}~\bibnamefont{Harvey-Collard}}, \bibinfo{author}{\bibfnamefont{G.}~\bibnamefont{Salis}}, \bibinfo{author}{\bibfnamefont{A.}~\bibnamefont{Fuhrer}}, \bibnamefont{et~al.}, \emph{\bibinfo{title}{Impact of interface traps on charge noise, mobility and percolation density in {Ge}/{SiGe} heterostructures}} (\bibinfo{year}{2023}), \bibinfo{note}{arXiv:2310.05902 [cond-mat]}, \urlprefix\url{http://arxiv.org/abs/2310.05902}.

\bibitem[{\citenamefont{Ruggiero et~al.}()\citenamefont{Ruggiero, Nigro, Zardo, and Hofmann}}]{ruggiero_backgate_2024}
\bibinfo{author}{\bibfnamefont{L.}~\bibnamefont{Ruggiero}}, \bibinfo{author}{\bibfnamefont{A.}~\bibnamefont{Nigro}}, \bibinfo{author}{\bibfnamefont{I.}~\bibnamefont{Zardo}}, \bibnamefont{and} \bibinfo{author}{\bibfnamefont{A.}~\bibnamefont{Hofmann}}, \emph{\bibinfo{title}{A backgate for enhanced tunability of holes in planar germanium}}, \eprint{2407.15725 [cond-mat, physics:quant-ph]}, \urlprefix\url{http://arxiv.org/abs/2407.15725}.

\bibitem[{\citenamefont{Yoneda et~al.}(2018)\citenamefont{Yoneda, Takeda, Otsuka, Nakajima, Delbecq, Allison, Honda, Kodera, Oda, Hoshi et~al.}}]{yoneda_quantum-dot_2018}
\bibinfo{author}{\bibfnamefont{J.}~\bibnamefont{Yoneda}}, \bibinfo{author}{\bibfnamefont{K.}~\bibnamefont{Takeda}}, \bibinfo{author}{\bibfnamefont{T.}~\bibnamefont{Otsuka}}, \bibinfo{author}{\bibfnamefont{T.}~\bibnamefont{Nakajima}}, \bibinfo{author}{\bibfnamefont{M.~R.} \bibnamefont{Delbecq}}, \bibinfo{author}{\bibfnamefont{G.}~\bibnamefont{Allison}}, \bibinfo{author}{\bibfnamefont{T.}~\bibnamefont{Honda}}, \bibinfo{author}{\bibfnamefont{T.}~\bibnamefont{Kodera}}, \bibinfo{author}{\bibfnamefont{S.}~\bibnamefont{Oda}}, \bibinfo{author}{\bibfnamefont{Y.}~\bibnamefont{Hoshi}}, \bibnamefont{et~al.}, \bibinfo{journal}{Nature Nanotechnology} \textbf{\bibinfo{volume}{13}}, \bibinfo{pages}{102} (\bibinfo{year}{2018}), ISSN \bibinfo{issn}{1748-3395}, \urlprefix\url{https://www.nature.com/articles/s41565-017-0014-x}.

\bibitem[{\citenamefont{Kuhlmann et~al.}(2013)\citenamefont{Kuhlmann, Houel, Ludwig, Greuter, Reuter, Wieck, Poggio, and Warburton}}]{kuhlmann_charge_2013}
\bibinfo{author}{\bibfnamefont{A.~V.} \bibnamefont{Kuhlmann}}, \bibinfo{author}{\bibfnamefont{J.}~\bibnamefont{Houel}}, \bibinfo{author}{\bibfnamefont{A.}~\bibnamefont{Ludwig}}, \bibinfo{author}{\bibfnamefont{L.}~\bibnamefont{Greuter}}, \bibinfo{author}{\bibfnamefont{D.}~\bibnamefont{Reuter}}, \bibinfo{author}{\bibfnamefont{A.~D.} \bibnamefont{Wieck}}, \bibinfo{author}{\bibfnamefont{M.}~\bibnamefont{Poggio}}, \bibnamefont{and} \bibinfo{author}{\bibfnamefont{R.~J.} \bibnamefont{Warburton}}, \bibinfo{journal}{Nature Physics} \textbf{\bibinfo{volume}{9}}, \bibinfo{pages}{570} (\bibinfo{year}{2013}), ISSN \bibinfo{issn}{1745-2481}, \urlprefix\url{https://www.nature.com/articles/nphys2688}.

\bibitem[{\citenamefont{Bermeister et~al.}(2014)\citenamefont{Bermeister, Keith, and Culcer}}]{bermeister_charge_2014}
\bibinfo{author}{\bibfnamefont{A.}~\bibnamefont{Bermeister}}, \bibinfo{author}{\bibfnamefont{D.}~\bibnamefont{Keith}}, \bibnamefont{and} \bibinfo{author}{\bibfnamefont{D.}~\bibnamefont{Culcer}}, \bibinfo{journal}{Applied Physics Letters} \textbf{\bibinfo{volume}{105}}, \bibinfo{pages}{192102} (\bibinfo{year}{2014}), ISSN \bibinfo{issn}{0003-6951, 1077-3118}, \urlprefix\url{https://pubs.aip.org/apl/article/105/19/192102/596637/Charge-noise-spin-orbit-coupling-and-dephasing-of}.

\bibitem[{\citenamefont{Nigro et~al.}(2024{\natexlab{b}})\citenamefont{Nigro, Jutzi, Oppliger, De~Palma, Olsen, Ruiz-Caridad, Gadea, Scarlino, Zardo, and Hofmann}}]{nigro_demonstration_2024}
\bibinfo{author}{\bibfnamefont{A.}~\bibnamefont{Nigro}}, \bibinfo{author}{\bibfnamefont{E.}~\bibnamefont{Jutzi}}, \bibinfo{author}{\bibfnamefont{F.}~\bibnamefont{Oppliger}}, \bibinfo{author}{\bibfnamefont{F.}~\bibnamefont{De~Palma}}, \bibinfo{author}{\bibfnamefont{C.}~\bibnamefont{Olsen}}, \bibinfo{author}{\bibfnamefont{A.}~\bibnamefont{Ruiz-Caridad}}, \bibinfo{author}{\bibfnamefont{G.}~\bibnamefont{Gadea}}, \bibinfo{author}{\bibfnamefont{P.}~\bibnamefont{Scarlino}}, \bibinfo{author}{\bibfnamefont{I.}~\bibnamefont{Zardo}}, \bibnamefont{and} \bibinfo{author}{\bibfnamefont{A.}~\bibnamefont{Hofmann}}, \bibinfo{journal}{ACS Applied Electronic Materials} \textbf{\bibinfo{volume}{6}}, \bibinfo{pages}{5094} (\bibinfo{year}{2024}{\natexlab{b}}), ISSN \bibinfo{issn}{2637-6113, 2637-6113}, \urlprefix\url{https://pubs.acs.org/doi/10.1021/acsaelm.4c00654}.

\bibitem[{\citenamefont{Sun et~al.}(2006)\citenamefont{Sun, Sun, Liu, Lee, Peterson, and Pianetta}}]{sun_surface_2006}
\bibinfo{author}{\bibfnamefont{S.}~\bibnamefont{Sun}}, \bibinfo{author}{\bibfnamefont{Y.}~\bibnamefont{Sun}}, \bibinfo{author}{\bibfnamefont{Z.}~\bibnamefont{Liu}}, \bibinfo{author}{\bibfnamefont{D.-I.} \bibnamefont{Lee}}, \bibinfo{author}{\bibfnamefont{S.}~\bibnamefont{Peterson}}, \bibnamefont{and} \bibinfo{author}{\bibfnamefont{P.}~\bibnamefont{Pianetta}}, \bibinfo{journal}{Applied Physics Letters} \textbf{\bibinfo{volume}{88}}, \bibinfo{pages}{021903} (\bibinfo{year}{2006}), ISSN \bibinfo{issn}{0003-6951, 1077-3118}, \urlprefix\url{https://pubs.aip.org/apl/article/88/2/021903/921491/Surface-termination-and-roughness-of-Ge-100}.

\bibitem[{\citenamefont{Ponath et~al.}(2013)\citenamefont{Ponath, Posadas, Hatch, and Demkov}}]{ponath_preparation_2013}
\bibinfo{author}{\bibfnamefont{P.}~\bibnamefont{Ponath}}, \bibinfo{author}{\bibfnamefont{A.~B.} \bibnamefont{Posadas}}, \bibinfo{author}{\bibfnamefont{R.~C.} \bibnamefont{Hatch}}, \bibnamefont{and} \bibinfo{author}{\bibfnamefont{A.~A.} \bibnamefont{Demkov}}, \bibinfo{journal}{Journal of Vacuum Science \& Technology B, Nanotechnology and Microelectronics: Materials, Processing, Measurement, and Phenomena} \textbf{\bibinfo{volume}{31}}, \bibinfo{pages}{031201} (\bibinfo{year}{2013}), ISSN \bibinfo{issn}{2166-2746, 2166-2754}, \urlprefix\url{https://pubs.aip.org/jvb/article/31/3/031201/102398/Preparation-of-a-clean-Ge-001-surface-using-oxygen}.

\bibitem[{\citenamefont{Ponath et~al.}(2017)\citenamefont{Ponath, Posadas, and Demkov}}]{ponath_ge001_2017}
\bibinfo{author}{\bibfnamefont{P.}~\bibnamefont{Ponath}}, \bibinfo{author}{\bibfnamefont{A.~B.} \bibnamefont{Posadas}}, \bibnamefont{and} \bibinfo{author}{\bibfnamefont{A.~A.} \bibnamefont{Demkov}}, \bibinfo{journal}{Applied Physics Reviews} \textbf{\bibinfo{volume}{4}}, \bibinfo{pages}{021308} (\bibinfo{year}{2017}), ISSN \bibinfo{issn}{1931-9401}, \urlprefix\url{https://pubs.aip.org/apr/article/4/2/021308/279713/Ge-001-surface-cleaning-methods-for-device}.

\bibitem[{\citenamefont{Groner et~al.}(2002)\citenamefont{Groner, Elam, Fabreguette, and George}}]{groner_electrical_2002}
\bibinfo{author}{\bibfnamefont{M.~D.} \bibnamefont{Groner}}, \bibinfo{author}{\bibfnamefont{J.~W.} \bibnamefont{Elam}}, \bibinfo{author}{\bibfnamefont{F.~H.} \bibnamefont{Fabreguette}}, \bibnamefont{and} \bibinfo{author}{\bibfnamefont{S.~M.} \bibnamefont{George}}, \bibinfo{journal}{Thin Solid Films} \textbf{\bibinfo{volume}{413}}, \bibinfo{pages}{186} (\bibinfo{year}{2002}), ISSN \bibinfo{issn}{0040-6090}, \urlprefix\url{https://www.sciencedirect.com/science/article/pii/S0040609002004388}.

\bibitem[{\citenamefont{Groner et~al.}(2004)\citenamefont{Groner, Fabreguette, Elam, and George}}]{groner_low-temperature_2004}
\bibinfo{author}{\bibfnamefont{M.~D.} \bibnamefont{Groner}}, \bibinfo{author}{\bibfnamefont{F.~H.} \bibnamefont{Fabreguette}}, \bibinfo{author}{\bibfnamefont{J.~W.} \bibnamefont{Elam}}, \bibnamefont{and} \bibinfo{author}{\bibfnamefont{S.~M.} \bibnamefont{George}}, \bibinfo{journal}{Chemistry of Materials} \textbf{\bibinfo{volume}{16}}, \bibinfo{pages}{639} (\bibinfo{year}{2004}), ISSN \bibinfo{issn}{0897-4756, 1520-5002}, \urlprefix\url{https://pubs.acs.org/doi/10.1021/cm0304546}.

\bibitem[{\citenamefont{Sioncke et~al.}(2009)\citenamefont{Sioncke, Delabie, Brammertz, Conard, Franquet, Caymax, Urbanzcyk, Heyns, Meuris, Van~Hemmen et~al.}}]{sioncke_thermal_2009}
\bibinfo{author}{\bibfnamefont{S.}~\bibnamefont{Sioncke}}, \bibinfo{author}{\bibfnamefont{A.}~\bibnamefont{Delabie}}, \bibinfo{author}{\bibfnamefont{G.}~\bibnamefont{Brammertz}}, \bibinfo{author}{\bibfnamefont{T.}~\bibnamefont{Conard}}, \bibinfo{author}{\bibfnamefont{A.}~\bibnamefont{Franquet}}, \bibinfo{author}{\bibfnamefont{M.}~\bibnamefont{Caymax}}, \bibinfo{author}{\bibfnamefont{A.}~\bibnamefont{Urbanzcyk}}, \bibinfo{author}{\bibfnamefont{M.}~\bibnamefont{Heyns}}, \bibinfo{author}{\bibfnamefont{M.}~\bibnamefont{Meuris}}, \bibinfo{author}{\bibfnamefont{J.~L.} \bibnamefont{Van~Hemmen}}, \bibnamefont{et~al.}, \bibinfo{journal}{Journal of The Electrochemical Society} \textbf{\bibinfo{volume}{156}}, \bibinfo{pages}{H255} (\bibinfo{year}{2009}), ISSN \bibinfo{issn}{00134651}, \urlprefix\url{https://iopscience.iop.org/article/10.1149/1.3076143}.

\bibitem[{\citenamefont{Kim et~al.}(2017)\citenamefont{Kim, Tyryshkin, and Lyon}}]{kim_annealing_2017}
\bibinfo{author}{\bibfnamefont{J.-S.} \bibnamefont{Kim}}, \bibinfo{author}{\bibfnamefont{A.~M.} \bibnamefont{Tyryshkin}}, \bibnamefont{and} \bibinfo{author}{\bibfnamefont{S.~A.} \bibnamefont{Lyon}}, \bibinfo{journal}{Applied Physics Letters} \textbf{\bibinfo{volume}{110}}, \bibinfo{pages}{123505} (\bibinfo{year}{2017}), ISSN \bibinfo{issn}{0003-6951, 1077-3118}, \urlprefix\url{https://pubs.aip.org/apl/article/110/12/123505/33194/Annealing-shallow-Si-SiO2-interface-traps-in}.

\bibitem[{\citenamefont{Kim et~al.}(2022)\citenamefont{Kim, Lee, Jo, Seo, Yoo, and Kim}}]{kim_influence_2022}
\bibinfo{author}{\bibfnamefont{S.}~\bibnamefont{Kim}}, \bibinfo{author}{\bibfnamefont{S.-H.} \bibnamefont{Lee}}, \bibinfo{author}{\bibfnamefont{I.~H.} \bibnamefont{Jo}}, \bibinfo{author}{\bibfnamefont{J.}~\bibnamefont{Seo}}, \bibinfo{author}{\bibfnamefont{Y.-E.} \bibnamefont{Yoo}}, \bibnamefont{and} \bibinfo{author}{\bibfnamefont{J.~H.} \bibnamefont{Kim}}, \bibinfo{journal}{Scientific Reports} \textbf{\bibinfo{volume}{12}}, \bibinfo{pages}{5124} (\bibinfo{year}{2022}), ISSN \bibinfo{issn}{2045-2322}, \urlprefix\url{https://www.nature.com/articles/s41598-022-09054-7}.

\bibitem[{\citenamefont{Paghi et~al.}()\citenamefont{Paghi, Battisti, Tortorella, De~Simoni, and Giazotto}}]{paghi_cryogenic_2024}
\bibinfo{author}{\bibfnamefont{A.}~\bibnamefont{Paghi}}, \bibinfo{author}{\bibfnamefont{S.}~\bibnamefont{Battisti}}, \bibinfo{author}{\bibfnamefont{S.}~\bibnamefont{Tortorella}}, \bibinfo{author}{\bibfnamefont{G.}~\bibnamefont{De~Simoni}}, \bibnamefont{and} \bibinfo{author}{\bibfnamefont{F.}~\bibnamefont{Giazotto}}, \emph{\bibinfo{title}{Cryogenic behavior of high-permittivity gate dielectrics: The impact of the atomic layer deposition temperature and the lithography pattering method}}, \eprint{2407.04501 [cond-mat]}, \urlprefix\url{http://arxiv.org/abs/2407.04501}.

\bibitem[{\citenamefont{Dimoulas et~al.}(2006)\citenamefont{Dimoulas, Tsipas, Sotiropoulos, and Evangelou}}]{dimoulas_fermi-level_2006}
\bibinfo{author}{\bibfnamefont{A.}~\bibnamefont{Dimoulas}}, \bibinfo{author}{\bibfnamefont{P.}~\bibnamefont{Tsipas}}, \bibinfo{author}{\bibfnamefont{A.}~\bibnamefont{Sotiropoulos}}, \bibnamefont{and} \bibinfo{author}{\bibfnamefont{E.~K.} \bibnamefont{Evangelou}}, \bibinfo{journal}{Applied Physics Letters} \textbf{\bibinfo{volume}{89}}, \bibinfo{pages}{252110} (\bibinfo{year}{2006}), ISSN \bibinfo{issn}{0003-6951, 1077-3118}, \urlprefix\url{https://pubs.aip.org/apl/article/89/25/252110/921609/Fermi-level-pinning-and-charge-neutrality-level-in}.

\bibitem[{\citenamefont{Morita et~al.}(1990)\citenamefont{Morita, Ohmi, Hasegawa, Kawakami, and Ohwada}}]{morita_growth_1990}
\bibinfo{author}{\bibfnamefont{M.}~\bibnamefont{Morita}}, \bibinfo{author}{\bibfnamefont{T.}~\bibnamefont{Ohmi}}, \bibinfo{author}{\bibfnamefont{E.}~\bibnamefont{Hasegawa}}, \bibinfo{author}{\bibfnamefont{M.}~\bibnamefont{Kawakami}}, \bibnamefont{and} \bibinfo{author}{\bibfnamefont{M.}~\bibnamefont{Ohwada}}, \bibinfo{journal}{Journal of Applied Physics} \textbf{\bibinfo{volume}{68}}, \bibinfo{pages}{1272} (\bibinfo{year}{1990}), ISSN \bibinfo{issn}{0021-8979, 1089-7550}, \urlprefix\url{https://pubs.aip.org/jap/article/68/3/1272/18021/Growth-of-native-oxide-on-a-silicon-surface}.

\bibitem[{\citenamefont{Bohling and Sigmund}(2016)}]{bohling_self-limitation_2016}
\bibinfo{author}{\bibfnamefont{C.}~\bibnamefont{Bohling}} \bibnamefont{and} \bibinfo{author}{\bibfnamefont{W.}~\bibnamefont{Sigmund}}, \bibinfo{journal}{Silicon} \textbf{\bibinfo{volume}{8}}, \bibinfo{pages}{339} (\bibinfo{year}{2016}), ISSN \bibinfo{issn}{1876-9918}, \urlprefix\url{https://doi.org/10.1007/s12633-015-9366-8}.

\bibitem[{\citenamefont{Degli~Esposti et~al.}(2022)\citenamefont{Degli~Esposti, Paquelet~Wuetz, Fezzi, Lodari, Sammak, and Scappucci}}]{degli_esposti_wafer-scale_2022}
\bibinfo{author}{\bibfnamefont{D.}~\bibnamefont{Degli~Esposti}}, \bibinfo{author}{\bibfnamefont{B.}~\bibnamefont{Paquelet~Wuetz}}, \bibinfo{author}{\bibfnamefont{V.}~\bibnamefont{Fezzi}}, \bibinfo{author}{\bibfnamefont{M.}~\bibnamefont{Lodari}}, \bibinfo{author}{\bibfnamefont{A.}~\bibnamefont{Sammak}}, \bibnamefont{and} \bibinfo{author}{\bibfnamefont{G.}~\bibnamefont{Scappucci}}, \bibinfo{journal}{Applied Physics Letters} \textbf{\bibinfo{volume}{120}}, \bibinfo{pages}{184003} (\bibinfo{year}{2022}), ISSN \bibinfo{issn}{0003-6951, 1077-3118}, \urlprefix\url{https://pubs.aip.org/apl/article/120/18/184003/2833621/Wafer-scale-low-disorder-2DEG-in-28Si-SiGe-without}.

\bibitem[{\citenamefont{Xu et~al.}(2021)\citenamefont{Xu, Li, Zhang, Yan, Liu, Yuan, Ye, and Li}}]{xu_impact_2021}
\bibinfo{author}{\bibfnamefont{J.}~\bibnamefont{Xu}}, \bibinfo{author}{\bibfnamefont{S.}~\bibnamefont{Li}}, \bibinfo{author}{\bibfnamefont{W.}~\bibnamefont{Zhang}}, \bibinfo{author}{\bibfnamefont{S.}~\bibnamefont{Yan}}, \bibinfo{author}{\bibfnamefont{C.}~\bibnamefont{Liu}}, \bibinfo{author}{\bibfnamefont{X.}~\bibnamefont{Yuan}}, \bibinfo{author}{\bibfnamefont{X.}~\bibnamefont{Ye}}, \bibnamefont{and} \bibinfo{author}{\bibfnamefont{H.}~\bibnamefont{Li}}, \bibinfo{journal}{Applied Surface Science} \textbf{\bibinfo{volume}{544}}, \bibinfo{pages}{148889} (\bibinfo{year}{2021}), ISSN \bibinfo{issn}{01694332}, \urlprefix\url{https://linkinghub.elsevier.com/retrieve/pii/S0169433220336485}.

\bibitem[{\citenamefont{Benick et~al.}(2010)\citenamefont{Benick, Richter, Li, Grant, McIntosh, Ren, Weber, Hermle, and Glunz}}]{benick_effect_2010}
\bibinfo{author}{\bibfnamefont{J.}~\bibnamefont{Benick}}, \bibinfo{author}{\bibfnamefont{A.}~\bibnamefont{Richter}}, \bibinfo{author}{\bibfnamefont{T.-T.~A.} \bibnamefont{Li}}, \bibinfo{author}{\bibfnamefont{N.~E.} \bibnamefont{Grant}}, \bibinfo{author}{\bibfnamefont{K.~R.} \bibnamefont{McIntosh}}, \bibinfo{author}{\bibfnamefont{Y.}~\bibnamefont{Ren}}, \bibinfo{author}{\bibfnamefont{K.~J.} \bibnamefont{Weber}}, \bibinfo{author}{\bibfnamefont{M.}~\bibnamefont{Hermle}}, \bibnamefont{and} \bibinfo{author}{\bibfnamefont{S.~W.} \bibnamefont{Glunz}}, in \emph{\bibinfo{booktitle}{2010 35th {IEEE} {Photovoltaic} {Specialists} {Conference}}} (\bibinfo{publisher}{IEEE}, \bibinfo{address}{Honolulu, HI, USA}, \bibinfo{year}{2010}), pp. \bibinfo{pages}{000891--000896}, ISBN \bibinfo{isbn}{9781424458905 9781424458929}, \urlprefix\url{https://ieeexplore.ieee.org/document/5614148/}.

\bibitem[{\citenamefont{Weber et~al.}(2011)\citenamefont{Weber, Janotti, and Van De~Walle}}]{weber_native_2011}
\bibinfo{author}{\bibfnamefont{J.~R.} \bibnamefont{Weber}}, \bibinfo{author}{\bibfnamefont{A.}~\bibnamefont{Janotti}}, \bibnamefont{and} \bibinfo{author}{\bibfnamefont{C.~G.} \bibnamefont{Van De~Walle}}, \bibinfo{journal}{Journal of Applied Physics} \textbf{\bibinfo{volume}{109}}, \bibinfo{pages}{033715} (\bibinfo{year}{2011}), ISSN \bibinfo{issn}{0021-8979, 1089-7550}, \urlprefix\url{https://pubs.aip.org/jap/article/109/3/033715/925070/Native-defects-in-Al2O3-and-their-impact-on-III-V}.

\bibitem[{\citenamefont{Hensling et~al.}(2017)\citenamefont{Hensling, Xu, Gunkel, and Dittmann}}]{hensling_unraveling_2017}
\bibinfo{author}{\bibfnamefont{F.~V.~E.} \bibnamefont{Hensling}}, \bibinfo{author}{\bibfnamefont{C.}~\bibnamefont{Xu}}, \bibinfo{author}{\bibfnamefont{F.}~\bibnamefont{Gunkel}}, \bibnamefont{and} \bibinfo{author}{\bibfnamefont{R.}~\bibnamefont{Dittmann}}, \bibinfo{journal}{Scientific Reports} \textbf{\bibinfo{volume}{7}}, \bibinfo{pages}{39953} (\bibinfo{year}{2017}), ISSN \bibinfo{issn}{2045-2322}, \urlprefix\url{https://www.nature.com/articles/srep39953}.

\bibitem[{\citenamefont{Martinez et~al.}(2024)\citenamefont{Martinez, de~Franceschi, and Niquet}}]{martinez_variability_2024}
\bibinfo{author}{\bibfnamefont{B.}~\bibnamefont{Martinez}}, \bibinfo{author}{\bibfnamefont{S.}~\bibnamefont{de~Franceschi}}, \bibnamefont{and} \bibinfo{author}{\bibfnamefont{Y.-M.} \bibnamefont{Niquet}}, \emph{\bibinfo{title}{Variability mitigation in epitaxial-heterostructure-based spin qubit devices via gate layout optimization}} (\bibinfo{year}{2024}), \bibinfo{note}{arXiv:2402.18991 [cond-mat]}, \urlprefix\url{http://arxiv.org/abs/2402.18991}.

\bibitem[{\citenamefont{Cui et~al.}(2015)\citenamefont{Cui, Lee, Kim, Arefe, Huang, Lee, Chenet, Zhang, Wang, Ye et~al.}}]{cui_multi-terminal_2015}
\bibinfo{author}{\bibfnamefont{X.}~\bibnamefont{Cui}}, \bibinfo{author}{\bibfnamefont{G.-H.} \bibnamefont{Lee}}, \bibinfo{author}{\bibfnamefont{Y.~D.} \bibnamefont{Kim}}, \bibinfo{author}{\bibfnamefont{G.}~\bibnamefont{Arefe}}, \bibinfo{author}{\bibfnamefont{P.~Y.} \bibnamefont{Huang}}, \bibinfo{author}{\bibfnamefont{C.-H.} \bibnamefont{Lee}}, \bibinfo{author}{\bibfnamefont{D.~A.} \bibnamefont{Chenet}}, \bibinfo{author}{\bibfnamefont{X.}~\bibnamefont{Zhang}}, \bibinfo{author}{\bibfnamefont{L.}~\bibnamefont{Wang}}, \bibinfo{author}{\bibfnamefont{F.}~\bibnamefont{Ye}}, \bibnamefont{et~al.}, \bibinfo{journal}{Nature Nanotechnology} \textbf{\bibinfo{volume}{10}}, \bibinfo{pages}{534} (\bibinfo{year}{2015}), ISSN \bibinfo{issn}{1748-3395}, \urlprefix\url{https://www.nature.com/articles/nnano.2015.70}.

\bibitem[{\citenamefont{Monroe et~al.}(1993)\citenamefont{Monroe, Xie, Fitzgerald, Silverman, and Watson}}]{monroe_comparison_1993}
\bibinfo{author}{\bibfnamefont{D.}~\bibnamefont{Monroe}}, \bibinfo{author}{\bibfnamefont{Y.~H.} \bibnamefont{Xie}}, \bibinfo{author}{\bibfnamefont{E.~A.} \bibnamefont{Fitzgerald}}, \bibinfo{author}{\bibfnamefont{P.~J.} \bibnamefont{Silverman}}, \bibnamefont{and} \bibinfo{author}{\bibfnamefont{G.~P.} \bibnamefont{Watson}}, \bibinfo{journal}{Journal of Vacuum Science \& Technology B: Microelectronics and Nanometer Structures Processing, Measurement, and Phenomena} \textbf{\bibinfo{volume}{11}}, \bibinfo{pages}{1731} (\bibinfo{year}{1993}), ISSN \bibinfo{issn}{1071-1023, 1520-8567}, \urlprefix\url{https://pubs.aip.org/jvb/article/11/4/1731/1074186/Comparison-of-mobility-limiting-mechanisms-in-high}.

\bibitem[{\citenamefont{Tracy et~al.}(2009)\citenamefont{Tracy, Hwang, Eng, Ten~Eyck, Nordberg, Childs, Carroll, Lilly, and Das~Sarma}}]{tracy_observation_2009}
\bibinfo{author}{\bibfnamefont{L.~A.} \bibnamefont{Tracy}}, \bibinfo{author}{\bibfnamefont{E.~H.} \bibnamefont{Hwang}}, \bibinfo{author}{\bibfnamefont{K.}~\bibnamefont{Eng}}, \bibinfo{author}{\bibfnamefont{G.~A.} \bibnamefont{Ten~Eyck}}, \bibinfo{author}{\bibfnamefont{E.~P.} \bibnamefont{Nordberg}}, \bibinfo{author}{\bibfnamefont{K.}~\bibnamefont{Childs}}, \bibinfo{author}{\bibfnamefont{M.~S.} \bibnamefont{Carroll}}, \bibinfo{author}{\bibfnamefont{M.~P.} \bibnamefont{Lilly}}, \bibnamefont{and} \bibinfo{author}{\bibfnamefont{S.}~\bibnamefont{Das~Sarma}}, \bibinfo{journal}{Physical Review B} \textbf{\bibinfo{volume}{79}}, \bibinfo{pages}{235307} (\bibinfo{year}{2009}), \urlprefix\url{https://link.aps.org/doi/10.1103/PhysRevB.79.235307}.

\end{thebibliography}

\end{document}